\documentclass[12pt]{article}
\usepackage{amssymb,amsmath,epsfig}

\begin{document}

\title{\bf Dynamics of Shearfree Dissipative Collapse in $f(G)$ Gravity}

\author{Muhammad SHARIF \thanks{msharif.math@pu.edu.pk} and Ghulam ABBAS \thanks{abbasg91@yahoo.com}
\\
Department of Mathematics, University of the Punjab,\\
Quaid-e-Azam Campus, Lahore-54590, Pakistan.}

\date{}
\maketitle
\begin{abstract}
In this paper, we study the dynamics of shearfree dissipative
gravitational collapse in modified Gauss-Bonnet theory of gravity -
so called $f(G)$ gravity. The field equations of $f(G)$ gravity are
applied to shearfree spherical interior geometry of the dissipative
star. We formulate the dynamical as well as transport equations and
then couple them to investigate the process of collapse. We conclude
that the gravitational force in this theory is much stronger as
compared to general relativity which indicates increase in the rate
of collapse. Further, the relation between the Weyl tensor and
matter components is established. This shows that for constant
$f(G)$ the vanishing of the Weyl scalar leads to the homogeneity in
energy density and vice versa.
\end{abstract}

\section{Introduction}

Recently, a lot of work has been done on alternative theories of
gravity for the identification of dark energy (DE) which is said to
be responsible for accelerated expansion of the universe.$^{1)}$
These theories are attractive due to their consistency with the
astrophysical observations and local gravitational
experiments.$^{2)}$ The simplest generalization to general
relativity (GR) is $f(R)$ gravity in which $f$ is an arbitrary
function of the Ricci scalar $R$. Although, it is the simplest
modification of GR but it is not possible generally to formulate
such $f(R)$ model which is consistent with the solar system tests.
Many conditions for the validity of $f(R)$ models have been
proposed.$^{3)}$

Another modified theory of gravity "the modified Gauss-Bonnet
theory" was proposed by several authors.$^{4)-6)}$ In this theory,
GR is modified by introducing some arbitrary function $f(G)$ in
Einstein-Hilbert action, where
$G=R^2-4R^{\mu\nu}R_{\mu\nu}+R^{\mu\nu\gamma\delta}R_{\mu\nu\gamma\delta}$
is the Gauss-Bonnet invariant. This modification (consistent with
observational constraints) is related to string-inspired dilaton
theory.$^{7)}$ Cognola et al.$^{8)}$ found that such model can
produce transition form matter dominated era to accelerated phase.

The dynamics of the dissipative shearfree gravitational collapse is
an important issue. In GR, this problem was initially formulated
many years ago by Misner and Sharp$^{9)}$ for adiabatic collapse and
by Misner$^{10)}$ for non-adiabatic collapse. The dissipative
process during the collapse of massive star is important due to the
energy loss during formation of neutron star or black hole.$^{11)}$
It is shown$^{12),13)}$ that gravitational collapse is a dissipative
process, so the effects of the dissipation must be included in the
study of collapse. Herrera and Santos$^{14)}$ discussed the dynamics
of shearfree anisotropic radiating fluid. Chan$^{15)}$ studied the
realistic model of radiating star with shear viscosity.

Herrera$^{16)}$ explored the inertia of heat and its role in the
dynamics of dissipative collapse. Herrera et al.$^{17)}$ also
derived the dynamical equations by including dissipation in the form
of heat flow, radiation, shear and bulk viscosity and then coupled
with causal transport equation. They investigated that shearfree
condition for quasi-static (slowly evolving) non-dissipative systems
in Newtonian limit leads to homologous collapse. Di Prisco et
al.$^{18)}$ investigated the dynamics of spherically symmetric
charged viscous non-adiabatic gravitational collapse. Sharif and his
collaborators $^{19)-25)}$ have discussed the dynamics of charged
and uncharged viscous dissipative gravitational collapse in GR as
well as in $f(R)$ gravity.

In this paper, we discuss the dynamics of the dissipative shearfree
gravitational collapse in $f(G)$ gravity. We use the field equations
of $f(G)$ gravity$^{26)}$ developed by using covariant gauge
invariant (CGI) perturbation approach with $3+1$ formalism. The
collapsing matter is taken as dissipative isotropic fluid bounded by
the shearfree interior geometry. This spacetime is matched with the
Vaidya exterior solution by using Darmois junction conditions
$^{27)}$. The plan of the paper is as follows: In the next section,
we describe the field equations in $f(G)$ gravity and matching
conditions. We formulate the dynamical as well as transport
equations and then couple them in section \textbf{3}. In section
\textbf{4}, we discuss the relation between the Weyl tensor and
matter variables. The last section provides discussion of the
results.

\section{Field Equations in Modified Gauss-Bonnet Gravity}

The Einstein-Hilbert action for the modified Gauss-Bonnet gravity is
\begin{equation}\label{1}
S=\int d^4x\sqrt{-g}\left(\frac{R+f(G)}{2\kappa}+\mathcal{L}_m
\right),
\end{equation}
where $g$ is the determinant of the metric tensor $g_{\mu\nu},~R$ is
the Ricci scalar, $f(G)$ is an arbitrary function of Gauss Bonnet
invariant $G,~\mathcal{L}_m$ is the matter Lagrangian and $\kappa$
is coupling constant. The variation of this action with respect to
the metric tensor gives the field equations
\begin{eqnarray}\label{2}
G_{\mu\nu}&=&\kappa
T_{\mu\nu}+\frac{1}{2}g_{\mu\nu}f-2FRR_{\mu\nu}+4FR^{\gamma}_{\mu}R_{\nu\gamma}
-2FR_{\mu\gamma\delta\omega}R^{\gamma\delta\omega}_\nu\nonumber\\
&-&4FR_{\mu\gamma\delta\nu}R^{\gamma\delta}+2R\nabla_{\mu}\nabla_\nu
F-2g_{\mu\nu}\nabla^2F-4R^{\gamma}_{\mu}\nabla_{\nu}\nabla{\gamma}F\nonumber\\
&-&4R^{\gamma}_{\nu}\nabla_{\mu}\nabla{\gamma}F
+4R_{\mu\nu}\nabla^2F+4g_{\mu\nu}R^{\gamma\delta}\nabla_{\gamma}\nabla_{\delta}F
-4R_{\mu\gamma\nu\delta}\nabla^{\gamma}\nabla^{\delta}F,
\end{eqnarray}
where $F={\partial{f(G)}}/{\partial{G}}$. The energy-momentum tensor
for dissipative fluid is
\begin{equation}\label{3}
T_{\mu\nu}=({\rho}+p)V_{\mu}V_{\nu}-pg_{\mu\nu}+q_\mu V_\nu +q_\nu
V_\mu,
\end{equation}
where $\rho,~p,~V_\mu$ and $~q_\mu$ are density, pressure, four
velocity and radial heat flux, respectively.

In general, it is very difficult to handle the field equations. We
use a simplified form of these equations derived by Li et
al.$^{26)}$ through CGI perturbation using $3+1$ formalism and the
effective energy-momentum approach. Since there are nonlinear terms
in $R_{\mu\nu}$ and $R_{\mu\nu\lambda\sigma}$ which are expressed in
terms of dynamical quantities, so the field equations and
Gauss-Bonnet invariant are simplified as follows
\begin{equation}\label{4}
G_{\mu\nu}=\kappa(T_{\mu\nu}+T^{G}_{\mu\nu}),
\end{equation}
where $T^G_{\mu\nu}$ is the Gauss-Bonnet correction term. Using CGI
approach, the components of $T^G_{\mu\nu}$ are evaluated in terms of
the dynamical quantities as$^{26)}$
\begin{eqnarray}\label{5}
{\rho}^G&=&\frac{1}{\kappa}\left(\frac{1}{2}(f-FG)+\frac{2}{3}(\Omega-3\Psi)
(\dot{F}\Theta+{\tilde{\nabla}}^2F)\right),\\\label{6}
{-p}^G&=&\frac{1}{\kappa}\left(\frac{1}{2}(f-FG)+\frac{2}{3}(\Omega-3\Psi)\ddot{F}-
\frac{8}{9}\Omega(\Theta
\dot{F}+{\tilde{\nabla}}^2F)\right),\\\label{7}
q^G_{\mu}&=&\frac{1}{2}\left(-\frac{2}{3}(\Omega-3\Psi)(\tilde{\nabla}_\mu
\dot{F}-\frac{1}{3}\Theta\tilde{\nabla}_\mu{F})+\frac{4}{3}\dot{F}\Theta\zeta_\mu\right).
\end{eqnarray}
Here, $\tilde{\nabla}$ is spatial covariant derivative,
$\Theta=V^{\mu}_{~;\mu}$ is expansion scalar and the quantities
$\Omega,~\Psi,~\zeta_\mu$ are expressed in terms of dynamical
quantities as follows
\begin{eqnarray}\label{8}
\Omega&=&-(\dot{\Theta}+\frac{1}{3}\Theta^2-\tilde{\nabla}^{\mu}A_\mu),\\\label{9}
\Psi&=&-\frac{1}{3}(\dot{\Theta}+\Theta^2+\tilde{R}-\tilde{\nabla}^{\mu}A_\mu),\\\label{10}
\zeta_\mu&=&-\frac{2\tilde{\nabla}_\mu \Theta}{3}+\tilde{\nabla}^\nu
\sigma_{\mu\nu}+\tilde{\nabla}^\nu \omega_{\mu\nu},
\end{eqnarray}
where $\tilde{R}$ is the Ricci scalar of $3D$ spatial spherical
surface, $\sigma_{\mu\nu}=V_{(\mu;\nu)}-\dot{V}_{(\mu}V_{\nu)}-
\frac{1}{3}\Theta h_{\mu\nu}$ (where $h_{\mu\nu}=g_{\mu\nu}-V_{\mu}
V_\nu$) is the shear tensor, $A_{\mu}=V_{\mu;\nu}V^{\nu}$ is four
acceleration and
$\omega_{\mu\nu}=V_{[\mu;\nu]}+\dot{V}_{[\mu}V_{\nu]}$ is the
vorticity tensor. The Gauss-Bonnet invariant $G$ in CGI approach is
\begin{equation}\label{11}
G=2(\frac{1}{3}R^2-{R}^{\mu\nu}R_{\mu\nu}),
\end{equation}
where
\begin{eqnarray*}\label{12}
&R&=-2\dot{\Theta}-\frac{4}{3}\Theta^2+2\tilde{\nabla}^{\mu}A_\mu-\tilde{R},\\\label{13}
&R^{\mu\nu}R_{\mu\nu}&=\frac{4}{3}(\dot{\Theta}^2+\dot{\Theta}\Theta^2
+\frac{1}{3}\theta^4)+\frac{2}{3}(\dot{\Theta}+\Theta^2)\tilde{{R}}
-\frac{8}{3}(\dot{\Theta}+\frac{1}{2}\Theta^2)\tilde{\nabla}^{\mu}A_\mu.\nonumber
\end{eqnarray*}

Now we assume that the fluid (\ref{3}) is bounded by the spherical
boundary $\Sigma$. The metric interior to $\Sigma$ is assumed to be
comoving and shearfree in the following form
\begin{equation}\label{14}
ds^2=X^2dt^2-Y^2(dr^2+r^2d\theta^2+r^2\sin^2\theta d\phi^2),
\end{equation}
where $X$ and $Y$ are functions of $t$ and $r$. The corresponding
four velocity and heat flux take the form
\begin{equation}\label{15}
V^\mu=X^{-1}{\delta^\mu}_{0},\quad q^\mu=q{\delta^\mu}_{1},
\end{equation}
where $V^{\mu}q_{\mu}=0.$ The expansion scalar for the fluid sphere
is $\Theta=\frac{3\dot{Y}}{YX}$. Using Eqs.(\ref{3}), (\ref{4}),
(\ref{14}) and (\ref{15}), the field equations yield
\begin{eqnarray}\label{16}
8{\pi}(\rho+\rho^G)X^{2}&=&3\left(\frac{\dot{Y}}{Y}\right)^2
-\left(\frac{X}{Y}\right)^{2}\left(2\frac{Y''}{Y}
+\left(\frac{Y'}{Y}\right)^2+\frac{4Y'}{Y r}\right),\\\label{17}
8{\pi}(q+q^G)XY^2&=&{2}
\left(\frac{\dot{Y'}}{Y}-\frac{\dot{Y}Y'}{Y^2}-\frac{\dot{Y}X'}{YX}\right),\\\label{18}
8{\pi}(p+p^G) Y^{2}
&=&\left(\frac{Y'}{Y}\right)^{2}+\frac{2}{r}\left(\frac{Y'}{Y}+\frac{X'}{X}\right)+2\frac{X'Y'}{XY}
-\left(\frac{Y}{X}\right)^2\nonumber\\
&\times&\left(2\frac{\ddot{Y}}{Y}+\left(\frac{\dot{Y}}{Y}\right)^2
-2\frac{\dot{X}\dot{Y}}{XY}\right),\\\label{19} 8{\pi}(p+p^G)rY^{2}
&=&r^2\left(\frac{Y''}{Y}-\left(\frac{Y'}{Y}\right)^2+\frac{1}{r}\left(\frac{Y'}
{Y}+\frac{X'}{X}\right)+\frac{X''}{X}\right)\nonumber\\
&\times&\left(2\frac{\ddot{Y}}{Y}+\left(\frac{\dot{Y}}{Y}\right)^2-2\frac{\dot{X}\dot{Y}}{XY}\right),
\end{eqnarray}
where $\rho^G,~p^G$ and $q^G$ correspond to Gauss-Bonnet
contribution to GR. Here dot and prime indicate derivative with
respect to time and radial coordinate, respectively. Making use of
Eqs.(\ref{5})-(\ref{7}), these turn out to be
\begin{eqnarray}\label{20}
\kappa\rho^G&=&\frac{1}{2}(f-GF)+\frac{1}{3Y^2}\left(\frac{6\dot{Y}}{X^2}+\left(\frac{Y'}{Y}\right)^2
+\frac{4Y'}{Yr}+2\frac{Y''}{Y}\right)\nonumber\\
&\times&\left(\frac{3\dot{F}\dot{Y}}{XY}-\tilde{\nabla}^2F\right),
\end{eqnarray}
\begin{eqnarray}\label{21}
\kappa
p^G&=&\frac{1}{2}(GF-f)-\frac{1}{3Y^2}\left(\frac{6\dot{Y}}{X^2}
+\left(\frac{Y'}{Y}\right)^2+\frac{4Y'}{Yr}+2\frac{Y''}{Y}\right)\ddot{F}\nonumber\\
&-&\frac{8}{3}\left(\frac{\ddot{Y}}
{XY}-\frac{\dot{X}}{Y}-\frac{\dot{Y}}{XY}+\frac{\dot{Y}^2}{X^2Y^2}-\frac{1}{3Y^2}
\left(\frac{X''}{X}-\frac{X'^2}{X}+\frac{X'}{rX}\right)\right)\nonumber\\
&\times&\left(\frac{3\dot{F}\dot{Y}}{XY}-\tilde{\nabla}^2F\right),\\\label{22}
\kappa
q^G&=&\frac{4}{3Y^2}\left(\frac{6\dot{Y}}{X^2}+\left(\frac{Y'}{Y}\right)^2
+\frac{4Y'}{Yr}+2\frac{Y''}{Y}\right)\left(\frac{\dot{Y}F'}{YX}-\dot{F}'\right)\nonumber\\
&+&\frac{8\dot{F}\dot{X}}{YX}\left(\frac{\dot{Y}Y'}{XY^2}+\frac{\dot{Y}X'}{{X^2Y}}-\frac{\dot{Y'}}{XY}\right).
\end{eqnarray}
The Gauss-Bonnet invariant is found from Eq.(\ref{11}) as
\begin{eqnarray}\nonumber
G&=&\frac{2}{3}\left(-6
\left(\frac{\dot{Y}}{XY}\right)^{.}-12\frac{\dot{Y}^2}{(XY)^2}-\frac{2}{Y^2}
\left(\frac{X'Y'}{XY}-\left(\frac{X'}{X}\right)'\right)\right.\\\nonumber
&+&\left.4\frac{X'(Yr)'}{XY^3
r}-2\left(\frac{4Y'}{{Y^3r}}+\frac{Y''}{Y^3}-\frac{Y'^2}{{Y^4r}}\right)\right)^2-{8}\left
(\left(\frac{\dot{Y}}{(XY)}\right)^{.}\right)^2\nonumber\\
&+&\left(\frac{3\dot{Y}}{XY}\right)^2\left(\left(\frac{3\dot{Y}}{XY}\right)^{.}+\frac{1}{3}
\left(\frac{3\dot{Y}}{XY}\right)^2\right)+\frac{2}{3}
\left(\left(\frac{3\dot{Y}}{XY}\right)^{.}+\frac{1}{2}\left(\frac{3\dot{Y}}{XY}\right)^{2}
\right)\nonumber\\
&\times&\left(\frac{4Y'}{{Y^3r}}+\frac{Y''}{Y^3}-\frac{Y'^2}{{Y^4r}}+\frac{4}{Y^2}
\left(\frac{X'Y'}{XY}-\left(\frac{X}{X}\right)'\right)-\frac{X'Y'r}{XY^3r}\right).
\end{eqnarray}
It follows from Eq.(\ref{17}) that
\begin{equation}\label{23}
8{\pi}(q+q^G)Y^2=\frac{2\Theta '}{3}.
\end{equation}
If $q+q^G=0$, then $\Theta '=0$ and we have the condition for
homogenous collapse.

The Misner-Sharp mass becomes$^{9)}$
\begin{equation}\label{24}
m(r,t)=\frac{r^3}{2}\left(\frac{Y\dot{Y}^2}{X^2}-\frac{Y'^2}{Y}-\frac{2Y'}{r}\right).
\end{equation}
We assume that the exterior region to boundary surface $\Sigma$ is
described by the Vaidya spacetime which represents the geometry of
the radiating star given by
\begin{equation}\label{25}
{ds}^2=(1-\frac{M(\nu)}{\hat{R}})d\nu^2+2d\nu
d\hat{R}-\hat{R}^2(d\theta^2+\sin^2 \theta d \phi^2).
\end{equation}
Applying the same procedure as discussed in$^{19)-25)}$, we find
that the continuity of the first and second fundamental forms over
the boundary surface $\Sigma$ gives the continuity of the
gravitational masses across $\Sigma$ ($i.e.,~ m(r,t)=M(\nu)$) and
$p+p^G=^{\Sigma}(q+q^G)Y$. This implies that the effective pressure
is balanced by the effective radial heat flux on the boundary.

\section{Dynamical and Transport Equations}

In order to study the dynamics of the field equations, we use the
Misner-Sharp formalism$^{9)}$. Here velocity of the collapsing fluid
is given by
\begin{equation}\label{26}
U=rD_tY,
\end{equation}
where $U<0$ for radially inward motion of the fluid and
$D_t=\frac{1}{X}\frac{\partial}{\partial t}$. From the definition of
the mass function (\ref{24}), we can write
\begin{equation}\label{27}
\frac{(Yr)'}{Y}=\left(1+U^2+\frac{2m(r,t)}{rY}\right)^{\frac{1}{2}}=E.
\end{equation}
This is known as the energy of the collapsing fluid. The proper time
derivative of the mass function (\ref{24}) leads to
\begin{equation}\label{28a}
D_tm=r^3\left(\frac{Y\dot{Y}\ddot{Y}}{X^3}+\frac{1}{2}\left(\frac{\dot{Y}}{X}\right)^3
-\frac{Y{\dot{Y}}^2\dot{X}}{X^4}+\frac{1}{2}\frac{\dot{Y}Y'^2}{XY^2}
-r^3\frac{Y'\dot{Y'}}{YX}-\frac{\dot{Y}'}{rX}\right).
\end{equation}
Using Eqs.(\ref{17}), (\ref{18}), (\ref{25}) and (\ref{26}), this
can be written as
\begin{equation}\label{28}
D_t m=-4\pi \left[{(p+q^G)U}+(q+q^G)YE \right](rY)^2
\end{equation}
which give the variation of the fluid energy inside a sphere of
radius $rY$. Since $U<0$, so the first term on right side of this
equation increases the energy of the system while the negative sign
with second term implies the leaving of energy from the system.

We also define proper radial derivative
$D_R=\frac{1}{R'}\frac{\partial}{\partial r}$, where $R=rY$ is the
proper areal radius of the sphere. The proper radial derivative of
Eq.(\ref{24}) leads to
\begin{eqnarray}\nonumber
D_R{m} &=&\frac{Y}{(rY)'}
\left(-r^3\frac{Y'Y''}{Y^2}-r^2\frac{Y''}{Y}+\frac{r^3}{2}
\left(\frac{Y'}{Y}\right)^2-\frac{3r^2}{2}\left(\frac{Y'}{Y}\right)^2\right.\\\label{30}
&-&\left.2r\frac{Y'}{Y}-r^3\frac{X'\dot{Y}^2}{X^3}+\frac{r^3}{2}\frac{Y'\dot{Y}^2}{X^2Y}
+r^3\frac{\dot{Y'}\dot{Y}}{X^2}+\frac{3r^2}{2}\left(\frac{\dot{Y}}{X}\right)^2\right).
\end{eqnarray}
Using Eqs.(\ref{16}), (\ref{17}), (\ref{25}) and (\ref{26}) in
(\ref{30}), it follows that
\begin{equation}\label{31}
D_R m=4\pi\left((\rho+\rho^G)+(q+q^G)Y\frac{U}{E}\right)(rY)^2.
\end{equation}
This equation describes the variation in energy between two adjacent
layers of the fluid inside the spherical boundary. The Gauss-Bonnet
term effects the density and heat flux. Since heat flux is directed
outward due to $U<0$, so the Gauss-Bonnet term causes the system to
radiate away effectively.

The acceleration of the infalling matter inside the spherical
boundary is obtained by using Eqs.(\ref{18}), (\ref{24}) and
(\ref{26}) as
\begin{equation}\label{32}
D_tU=-\left(\frac{m}{(rY)^2}+4\pi(p+q^G)(rY)\right)+\frac{X'(rY)'}{XY^2}.
\end{equation}
The $r$ component of the energy-momentum conservation,
$(T^{1\beta}+T^{{G}{1\beta}})_{;\beta}=0$, implies that
\begin{eqnarray}\label{33}
\frac{1}{Y} (p+q^G)'+\frac{X'}{YX}(\rho+\rho^G+p+q^G)
+(\dot{q}+\dot{q}^G)\frac{Y}{X}+5(q+q^G)\frac{\dot{Y}}{X}=0.
\end{eqnarray}
Using the value of $\frac{X'}{X}$ from this equation in
Eq.(\ref{32}) with (\ref{24}) and (\ref{26}), we obtain the
dynamical equation
\begin{eqnarray}\label{34}
&&(\rho+\rho^G+p+q^G)D_tU=-(\rho+\rho^G+p+q^G)\left[m+4\pi(p+q^G)R^3\right]\frac{1}{R^2}\nonumber\\
&-&E^2\left[D_R(p+q^G)\right]-E[5Y(q+q^G)\frac{U}{R}+YD_t(q+q^G)].
\end{eqnarray}
This equation has the "Newtonian" form, i.e., $force=mass~density
\times acceleration$, where mass density $=\rho+\rho^G+p+q^G$. This
implies that Gauss-Bonnet term effects the mass density due to
higher curvature. The first square brackets on right side is the
gravitational force, whose Newtonian part is $m$ and relativistic
part is $p+q^G$, hence the Gauss-Bonnet term effects gravitational
force. The second term is the hydrodynamical force, it resists
against collapse because $D_R(p+q^G)<0$. The last square brackets
give the contribution of heat flux to the dynamics of the collapsing
system. Here the first term is positive ($U<0,~(q+q^G)>0$), implying
that outflow of heat flux reduces the rate of collapse by producing
radiations in the exterior of the collapsing sphere. The effects of
$D_t(q+q^G)$ will be explained below by introducing the heat
transport equation.

Now we discuss the transportation of heat during the shearfree
collapse of radiative fluid in $f(G)$ gravity. For this purpose, we
use the equation derived from Muller-Israel-Stewart phenomenological
theory for non-adiabatic fluids $^{28),29)}$. It is known that
Maxwell-Fourier law$^{30)}$ for the heat flux leads to diffusion
equation which indicates perturbation at very high speed. For the
relativistic non-adiabatic fluids, this is given by Eckart$^{31)}$
and Landau.$^{32)}$ To resolve this problem, many relativistic
theories$^{28)-30)}$ have been proposed. The common point of all
these theories is their validity for non-vanishing relaxation time.
These theories provide the heat transport equation which is
hyperbolic equation.

The transport equation for the heat flux in this case reads$^{14)}$
\begin{eqnarray}\label{1a}
\tau h^{\mu \nu}V^\lambda
\tilde{q}_{\nu;\lambda}+\tilde{q}^\mu=-\kappa h^{\mu
\nu}(T_{,\nu}+TA_\nu)+\frac{1}{2}\kappa T^2\left(\frac{\tau
V^\nu}{\kappa T^2}\right)_{;\nu}\tilde{q}^\mu,
\end{eqnarray}
where $h^{\mu\nu}$ is the projection tensor, $\tilde{q}=q+q^G$,
$\tau$ is relaxation time, $T$ is temperature and $\kappa$ is
thermal conductivity. For the interior spacetime, this equation
reduces to
\begin{eqnarray}\label{2a}
\tau \frac{\partial}{\partial
t}{[({q}+{q}^G)Y]}+(q+q^G)XY^2&=&-\kappa(TA)'-\frac{\kappa
T^2(q+q^G)Y^2}{2}(\frac{\tau}{\kappa T^2})\nonumber\\
&-&3\frac{\tau \dot{Y}Y(q+q^G)}{2}.
\end{eqnarray}
Using Eqs.(\ref{26}) and (\ref{31}), this implies
\begin{eqnarray}\label{3a}
&&YD_t(q+q^G)=-\kappa T \frac{D_t U}{\tau E}-\frac{\kappa
\acute{T}}{\tau Y}-\frac{Y(q+q^G)}{\tau}\left(1+\frac{\tau
U}{rY}\right)\nonumber\\
&-&\frac{\kappa T}{\tau E}[m+4\pi{(p+q^G)R^3}]R^{-2}-\frac{\kappa
T^2 (q+q^G)Y}{2XE}
\frac{\partial}{\partial t}\left(\frac{\tau}{\kappa T^2}\right)\nonumber\\
&-&\frac{3UY(q+q^G)}{2R}.
\end{eqnarray}

In order to see the effects of heat flux on the dynamics of
collapsing sphere in the modified Gauss-Bonnet gravity, we couple
the dynamical and heat transport equations given by Eqs.(\ref{34})
and (\ref{3a}). This coupling leads to
\begin{eqnarray}\label{3b}
&&(\rho+\rho^G+p+q^G)(1-\alpha)D_tU=F_{grav}(1-\alpha)+F_{hyd}+
\frac{E\kappa T'}{\tau Y}\nonumber\\
&+&\frac{E(q+q^G)Y}{\tau}-\frac{4E(q+q^G)YU}{R}+ \frac{\kappa T^2
(q+q^G)Y}{2XE} \frac{\partial}{\partial t}\left(\frac{\tau}{\kappa
T^2}\right)\nonumber\\
&+&\frac{3UY(q+q^G)}{2R}.
\end{eqnarray}
Here
\begin{eqnarray*}
F_{grav}&=&-(\rho+\rho^G+p+q^G){\left(m+4\pi(p+q^G)R^3\right)}\frac{1}{R^2},\\
F_{hyd}&=&-E^2[D_R(p+q^G)],\\
\alpha&=&\frac{\kappa T}{\tau(\rho+\rho^G+p+q^G)}.
\end{eqnarray*}
Now we analyze the effects of the Gauss-Bonnet term on the dynamics
of the collapsing radiative fluid. Since the presence of
Gauss-Bonnet term in $\alpha$ (thermal coefficient) effects the
value of $\alpha$ and the factor $(1-\alpha)$ being the the multiple
of $F_{grav}$ would effects the value of $F_{grav}$. In the
definition of $\alpha$, $\kappa$, $\tau$ and $T$ (as mentioned after
Eq.(\ref{1a})) are positive for a real dissipative gravitating
source $^{14)}$ and $(\rho+p)>0$, for collapsing fluid to satisfy
the weak energy condition. For $\rho^G$ and $q^G$, we have two cases
\begin{itemize}
\item When $\rho^G>0$ and $q^G>0$, then value of $\alpha$ in this case
will be less as compared to GR and value of $(1-\alpha)$ will be
larger (as compared to GR) and consequently gravitational force will
be stronger as compared to GR and rate of collapse will be
increased. For example if $f(G)=f_0$, a positive constant,
Eqs.(\ref{20})-(\ref{22}) yield $\rho^G=-p^G= \frac{f_0}{2}$ and
$q^G=0$ and we have $\rho^G=\frac{f_0}{2}>0$ and rate of collapse
will be increased.
\item When $\rho^G<0$ and $q^G<0$, then value of $\alpha$ in this case
will be larger as compared to GR and value of $(1-\alpha)$ will be
less (as compared to GR) and consequently gravitational force will
be weaker as compared to GR and rate of collapse will be decreased.
For example if $f(G)=f_0<0$ in above case, then we have
$\rho^G=\frac{f_0}{2}<0$ and rate of collapse will be decreased.
This can be verified for the more general $f(G)$ models.
\end{itemize}
If $\alpha$ approaches to a critical value of $1$ then the inertial
and gravitational forces of the system become zero and the system
will be in hydrostatic equilibrium. Notice that $\alpha$ appears in
the system due to its thermal conductivity and temperature. The
value of temperature for which $\alpha\rightarrow1$ is equivalent to
the expected amount of temperature that might be reached during the
radiative collapse in supernova explosion $^{33)}$. Thus $\alpha$
increases during the supernova explosion if the relaxation time
$\tau\neq0$. When $\alpha$ crosses the critical value, the
gravitational force plays the role of antigravity and expansion
would occur in the system.

\section{Relation Between Weyl Tensor and Matter Variables}

In this section, we establish a relation between the Weyl tensor and
density inhomogeneity which helps to extract some information about
the gravitational arrow of time. The Weyl tensor leads to the Weyl
scalar
$\mathcal{C}^{2}=C^{\mu\nu\lambda\sigma}C_{\mu\nu\lambda\sigma}$
which is given by
\begin{equation}\label{5a}
\mathcal{C}^{2}=\mathcal{R}-2\textit{R}^{\alpha\beta}\textit{R}_{\alpha\beta}
+\frac{1}{3}\textrm{R}^{2},
\end{equation}
where $\mathcal{R}=R^{\mu\nu\lambda\sigma}R_{\mu\nu\lambda\sigma}$
is the Kretchman scalar. Using the field equations in Eq.(\ref{3aa})
in appendix, Eq.(\ref{5a}) yields
\begin{equation}\label{6a}
{\epsilon}=m-\frac{4\pi}{3}R^2(\rho +\rho^G),
\end{equation}
where
\begin{equation}\label{7a}
{\epsilon}=\frac{\mathcal{C}}{{\sqrt{48}}}R^3.
\end{equation}
From Eqs.(\ref{31}) and (\ref{6a}), it follows that
\begin{eqnarray}\label{8a}
D_{R}{\epsilon}&=&4{\pi}R^2\left[(q+q^G)Y\frac{U}{E}-D_R(\rho+\rho^G)
\frac{R}{3}\right].
\end{eqnarray}

For $f(G)=f_0$, a constant, Eqs.(\ref{20})-(\ref{22}) yield
$\rho^G=-p^G= \frac{f_0}{2}$ and $q^G=0$, which implies that for
constant $f(G)$, we have $\Lambda$CDM, model. Assuming that the
fluid is non-dissipative, Eq.(\ref{8a}) yields
\begin{eqnarray}\label{9a}
D_{R}{\epsilon}+\frac{4{\pi}}{3}R^{3}D_{R}(\rho+\frac{f_0}{2})=0.
\end{eqnarray}
If we take $
D_{R}{\epsilon}=D_{R}\left(\frac{\mathcal{C}}{\sqrt{48}}R^{3}\right)=0$,
using regular axis condition, i.e., $R\neq0$, we have
$\mathcal{C}=0$. Thus Eq.(\ref{9a}) implies that $D_{R}\rho=0$. This
means that the conformal flatness condition produces homogeneity in
the energy density and vice versa. Equation (\ref{9a}) gives the
relationship between the energy density inhomogeneity and the Weyl
tenor for perfect fluid. This is an important expression for the
Penrose proposal about the gravitational arrow of time$^{34)}$. The
tidal forces associated with the Weyl tensor make the fluid more
inhomogeneous as the evolution occurs, indicating the sense of time.
We would like to mention here that such a relationship is no longer
valid if $f(G)$ is not a constant, like local anisotropy of the
pressure, dissipation and electric charge. In this case, we see from
Eq.(\ref{8a}) how $f(G)$ affects the link between the Weyl tensor
and energy density inhomogeneity, suggesting that $f(G)$ should
enter into the definition of gravitational arrow of time.

\section{Summary}

In this paper, we have considered a new framework in which an
arbitrary function $f(G)$ of Gauss-Bonnet invariant is introduced in
the Einstein-Hilbert action to account for the dynamics of the
shearfree dissipative gravitational collapse. The corresponding
field equations are used to study the shearfree dissipative interior
geometry of the star. The matching of the interior to the exterior
Vaidya geometry implies that the effective pressure is balanced by
the effective radial heat flux on the boundary. Using the Misner
definition, we have formulated the dynamical equation (\ref{34})
which shows that the Gauss-Bonnet correction increases (decreases)
the inertial mass leading to increase (decrease) the rate of
collapse.

To discuss the transportation of heat during the shearfree collapse
of radiative fluid, we have formulated heat transport equation.
Also, we have coupled the dynamical equation with the heat transport
equation to study the effects of radial heat flux on the dynamics of
collapsing fluid sphere. From this coupled equation ({\ref{3b}}), we
find that the Gauss-Bonnet term decreases (increases ) the value of
$\alpha$ for which $F_{grav}$ has larger (smaller) values which
increases the collapse of the system. Also, the equivalence
principle is valid in $f(G)$ gravity. We have explored that when
$\alpha$ exceeds the critical value, the gravitational force plays
the role of antigravity and reversal of collapse would occur in the
system. A model with bouncing behavior has been presented
numerically by Herrera et al.$^{35)}$ We would like to mention here
that the homogeneity in energy density and conformal flatness for
any metric are necessary and sufficient conditions for each other,
when $f(G)$ is constant and system under consideration is composed
of isotropic perfect fluid. This analysis can be extended for
generalized $f(G)$ model$^{36)}$ like $f(G)=C_1G+
C_2G^{\frac{1+\alpha}{4{\alpha}^2}}$, $C_1$, $C_2$ and $\alpha$ are
constants.

\section*{Appendix}

The interior metric (\ref{14}) has the following non-zero components
of the Riemann tensor
\begin{eqnarray}\nonumber
\textit{R}_{0101}&=&XX''+Y\ddot{Y}-\frac{X}{Y}X'Y'+\frac{Y}{X}\dot{X}\dot{Y},\nonumber\\
\textit{R}_{0202}&=&(Yr)^2\left(-\frac{\ddot{Y}}{Y}+\frac{\dot{X}\dot{Y}}{{XY}}
+\left(\frac{X}{Y}\right)^2\frac{X'}{X}(\frac{Y'}{Y}+\frac{1}{r})\right),\nonumber\\
\textit{R}_{0212}&=&(Yr)^2\left(-\frac{\dot{Y'}}{Y}+\frac{\dot{Y}{X'}}{{YX}}
+\frac{{X'}{\dot{Y}}}{YX}\right),\nonumber\\
\textit{R}_{1212}&=&(Yr)^2\left(\left(\frac{Y}{X}\right)^2\frac{\dot{Y}^2}{Y^2}
-\frac{Y''}{Y}+\frac{{Y'}^2}{Y^2}-\frac{Y'}{rY}\right),\nonumber\\
\textit{R}_{2323}&=&(Yr)^2\sin^2\theta\left(\left(\frac{r\dot{Y}}{Y}\right)^2
-\frac{\dot{rY'}^2}{Y^2}-2\frac{YrY'}{Y^2}\right), \nonumber\\
\textit{R}_{0303}&=&\sin^2\theta \textit{R}_{0202},\quad
\textit{R}_{0313}=\sin^2\theta\textit{R}_{0212},\quad
\textit{R}_{1313}=\sin^2\theta\textit{R}_{1212}.\label{1aa}
\end{eqnarray}
This implies that the Riemann tensor has five independent
components. The Kretchman scalar becomes
\begin{eqnarray}\nonumber
\mathcal{R}&=&4\left(\frac{1}{(XY)^4}{(R_{0101})^2}+\frac{2}{(XrY)^4}{(R_{0202})^2}
-\frac{4}{(XY^3r^2)^2}{(R_{0212})^2}\right.\\\label{2aa}
&+&\left.\frac{2}{(Y^2r)^4}{(R_{1212})^2}+\frac{1}{(Yr)^8\sin^2\theta}{(R_{2323})^2}\right).
\end{eqnarray}

The Riemann tensor components in terms of the Einstein tensor and
mass function can be written as
\begin{eqnarray}\nonumber
R_{0101}&=&(XY)^2\left(\frac{1}{2X^2}{G_{00}}-\frac{1}{2Y^2}{G_{11}}
+\frac{1}{(Yr)^2}{G_{22}}\right.-\left.\frac{2m}{(Yr)^3}\right),\\\nonumber
R_{0202}&=&(XYr)^2\left(\frac{1}{2Y^2}{G_{11}}+\frac{m}{(Yr)^3}\right),\\\nonumber
R_{0212}&=&\frac{(Yr)^2}{2}G_{01}, \\\nonumber
R_{1212}&=&(Y^2r)^2\left(\frac{1}{2X^2}{G_{00}}-\frac{m}{(Yr)^3}\right), \\
R_{2323}&=& 2mYr \sin^2\theta.
\end{eqnarray}
Inserting these in the Kretchman scalar (\ref{2aa}), it follows that
\begin{eqnarray}\label{3aa}
\mathcal{R}&=&\frac{48m^2}{(Yr)^6}-\frac{16m}{(Yr)^3}\left(\frac{G_{00}}
{{X^2}}-\frac{G_{11}}{{Y^2}}+\frac{G_{22}}{{(rY)^2}}\right)\nonumber\\
&-&4\left(\frac{G_{01}}{{XY}}\right)^2+3\left(\left(\frac{G_{00}}{{X^2}}
\right)^2+\left(\frac{G_{11}}{{Y^2}}\right)^2\right)+4\left(\frac{G_{22}}{{(rY)^2}}\right)^2\nonumber\\
&-&2\frac{G_{00}G_{11}}{X^2Y^2}+4\left(\frac{G_{00}}{X^2}-\frac{G_{11}}{Y^2}\right)\frac{G_{11}}{(rY)^2}.
\end{eqnarray}

\vspace{0.25cm}

{\bf Acknowledgment}

\vspace{0.25cm}

We would like to thank the Higher Education Commission, Islamabad,
Pakistan for its financial support through the {\it Indigenous Ph.D.
5000 Fellowship Program Batch-IV}.\\\\
 1) S. Nojiri and S. D. Odintsov: Phys. Rev. \textbf{D 75} (2006) 086005;
P. T. Sotiriou and S. Liberati: Ann. Phys. \textbf{322} (2007) 935;
P. T. Sotiriou and V. Faraoni: Rev. Mod. Phys. \textbf{82} (2010)
451; A. De Felice and S. Tsujikawa:
Living Rev. Rel. \textbf{13} (2010) 3.\\
2) G. R. Bengochea and R. Ferraro: Phys. Rev. \textbf{D 79} (2009)
124019; E. V. Linder: Phys. Rev. \textbf{D 81} (2010) 127301;
A. De Felice and S. Tsujikawa: Phys. Lett. \textbf{B 675} (2009) 1.\\
3) T. Harko, F. S. N. Lobo, S. Nojiri, and S. D. Odintsov: Phys.
Rev. \textbf{D 84} (2011) 024020; E. J. Copeland, M. Sami, and S.
Tsujikawa: Int. J. Mod. Phys. \textbf{D 15} (2006) 1753; S.
Capozzilo:
Int. J. Mod. Phys. \textbf{D 11} (2002) 483.\\
 4) S. Nojiri and S. D. Odintov:
 Phys. Lett. \textbf{B 631} (2005) 1.\\
5) S. Nojiri, S. D. Odintov and O. G. Gorbunova:
 J. Phys. \textbf{A 39} (2006) 6627.\\
6) G. Cognola, E. Elizalde, S. Nojiri, S. D. Odintov, and S.
Zerbini: Phys. Rev. \textbf{D 75} (2007) 086002.\\
7) M. Gasperini and G. Veneziano: Astropart. Phys.
\textbf{1} (1993) 317.\\
 8) G. Cognola, E. Elizalde, S. Nojiri,
S. D. Odintov, and S. Zerbini: Phys. Rev. \textbf{D 73} (2006) 084007.\\
9) C. W. Misner and D. H. Sharp: Phys. Rev.
\textbf{136} (1964) B571.\\
10) C. W. Misner: Phys. Rev. \textbf{137} (1965) B1360.\\
11) D. Kazanas and D. Schramm: \textit{Sources of Gravitational
Radiation} (Cambridge University Press, 1979).\\
12) L. Herrera, A. Di Prisco, J. Martin, J. Ospino, N. O. Santos,
and O. Troconis: Phys. Rev. \textbf{D 69} (2004) 084026.\\
13) A. Mitra: Phys. Rev. \textbf{D 74} (2006) 024010.\\
14) L. Herrera and N. O. Santos: Phys. Rev.
\textbf{D 70} (2004) 084004.\\
15) R. Chan: Astron. Astrophys. \textbf{368} (2001) 325.\\
16) L. Herrera: Int. J. Mod. Phys. \textbf{D 15} (2006) 2197.\\
17) L. Herrera, A. Di Prisco, E. Fuenmayor and O. Troconis: Int.
J. Mod. Phys. \textbf{D 18} (2009) 129.\\
18) A. Di Prisco, L. Herrera, G. Denmat, M. A. H. MacCallum, and
N. O. Santos: Phys. Rev. \textbf{D 80} (2009) 064031.\\
19) M. Sharif and G. Abbas: Astrophys. Space Sci. \textbf{335} (2011) 515.\\
20) M. Sharif and A. Siddiqa: Gen. Relativ. Gravity
\textbf{43} (2011) 37.\\
21) M. Sharif and S. Fatima: Gen. Relativ. Gravity
\textbf{43} (2011) 127.\\
 22) M. Sharif and Z. Rehmat: Gen. Relativ.
Gravity \textbf{42} (2010) 1795.\\
23) M. Sharif and H. R. Kausar: JCAP \textbf{07} (2011) 22.\\
24) M. Sharif and H. R. Kausar: Int. J. Mod. Phys.
\textbf{D 20} (2011) 2239.\\
25) M. Sharif and H. R. Kausar: Mod. Phys.
Lett. \textbf{A 25} (2010) 3299.\\
26) B. Li, D. J. Barrow, and F. D. Mota:
Phys. Rev. \textbf{D 76} (2007) 044027.\\
27) G. Darmois: \textit{Memorial des Sciences Mathematiques
}(Gautheir-Villars, Paris, 1927).\\
28) I. M$\ddot{u}$ller: Z. Physik \textbf{198} (1967) 329.\\
29) W. Israel and J. Stewart: Phys. Lett. \textbf{A 58} (1976) 2131.\\
30) D. Pav\'{e}n, D. Juo, and J. Casas-V\'{a}zques: Ann. Inst.
Henri Poincar\'{e} \textbf{A 36} (1982) 79.\\
31) C. Eckart: Phys. Rev. \textbf{58} (1940) 919.\\
32) L. Landau and E. Lifshitz: \textit{Fluid Mechanics} (Pergamon
Press, 1959).\\
33) J. Martinez: Phys. Rev. \textbf{D 53} (1996) 6921. \\
34) R. Penrose: \textit{General Relativity, An Einstein Centenary
Survey}(Cambridge University Press, 1979).\\
35) L. Herrera, A. Di Prisco, and W. Barreto: Phys. Rev. \textbf{D
73} (2006) 024008.\\
36) R. Myrzakulov, D. S$\acute{a}$ez-G$\acute{o}$emz and A. Tureanu:
 Gen. Relativ. Gravity \textbf{43} (2011) 1671.
\end{document}